\documentclass[sigconf]{acmart}
\setcopyright{none} 

\AtBeginDocument{%
  \providecommand\BibTeX{{%
    \normalfont B\kern-0.5em{\scshape i\kern-0.25em b}\kern-0.8em\TeX}}}

\renewcommand\footnotetextcopyrightpermission[1]{}

\begin{document}

\title{EmoG- Towards Emojifying Gmail Conversations}
\titlenote{To Appear in Proceedings of 54th Hawaii International Conference on System Sciences, HICSS 2021}

\author{Akhila Sri Manasa Venigalla \and Sridhar Chimalakonda}

\affiliation{%
   \institution{\textit{Research in Intelligent Software \& Human Analytics (RISHA) Lab}}
  \institution{Department of Computer Science and Engineering, \\Inidan Institute of Technology Tirupati}
  \city{Tirupati}
  \state{Andhra Pradesh}
    \country{India}}
\email{cs19d504@iittp.ac.in, ch@iittp.ac.in}


\begin{abstract}
  Emails are one of the most frequently used medium of communication in the present day across multiple domains including industry and educational institutions. Understanding sentiments being expressed in an email could have a considerable impact on the recipients' action or response to the email. However, it is difficult to interpret emotions of the sender from pure text in which emotions are not explicitly present. Researchers have tried to predict customer attrition by integrating emails in client-company environment with emotions. However, most of the existing works deal with static assessment of email emotions. Presenting sentiments of emails dynamically to the reader could help in understanding senders' emotion and as well have an impact on readers' action. Hence, in this paper, we present EmoG as a Google Chrome Extension which is intended to support university students. It augments emails with emojis based on the sentiment being conveyed in the email, which might also offer faster overview of email sentiments and act as tags that could help in automatic sorting and processing of emails. Currently, EmoG has been developed to support Gmail inbox on a Google Chrome browser, and could be extended to other inboxes and browsers with ease. We have conducted a user survey with 15 university students to understand the usefulness of EmoG and received positive feedback.
\end{abstract}



\maketitle
\pagestyle{plain}
\section{Introduction}
Emails are extensively used to exchange information among individuals and groups in various organizations, and include multiple levels of formality \cite{peterson2011email}. 
Among various CMCs, Emails are being frequently used in many organizations for official purposes such as reminders, distribution, and tracking of tasks, data stores and so on \cite{mark2016email}. Emails are also used in educational institutions for academic and administrative announcements to students and staff of the institution \cite{rashid2016technology}. They also act as a primary medium of communication between faculty and students to convey information about assignments, subjects, deadlines and so on, among student bodies to convey details about various events, discuss about multiple issues and so on, in many educational institutions \cite{couss2009attr}. It has been observed that email receivers try to perceive emotions of sender and that chances of miscommunicating emotions is considerably high \cite{byron2008carrying}. Adding emotions to emails might help in reducing misinterpretation of emails.

%

Pierce at al. have observed that few individuals find it comfortable conversing to others over text and that text messaging reduces societal anxiety \cite{pierce2009social}. 
However, it has also been noticed that many people find it easy to express their emotions through speech than with pure text \cite{holtzman2017emotional, colbert2016digital}. A study conducted by Sherman et al. reveals that individuals who use video chats or phone calls to communicate develop better bonding in comparison to individuals who communicate through text messages \cite{sherman2013effects}. The introduction of emoticons and emojis has facilitated expression of emotions through text, reducing the dependency on audio or face-to-face conversations to convey emotions. Using emojis in the text could help in conveying emotions better, as such visual cues could serve as motivation factors to encourage readers to read the email content \cite{ernst2018effects}. Considering the ability of emojis to represent sentiments and the wide usage of emails, inclusion of emojis in email texts might help in better representation of sentiments and might consequently improve readers' perception on emotions of the sender \cite{riordan2017emoji, coyle2019emoji}. However, it has been observed that emojis are scarcely used in emails that are communicated in educational or professional organizations. 

\begin{figure}
  \centering
    \includegraphics[width=\linewidth]{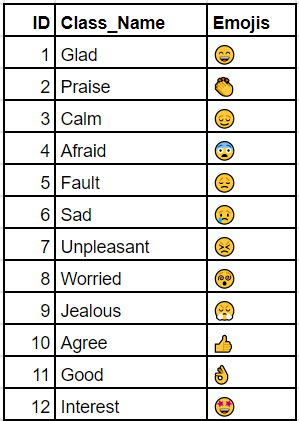}
    \caption{Twelve classes and corresponding emojis}
    \label{fig:classemo}
 \end{figure}

Marder et al. have performed studies that reveal the impact of using emojis in emails \cite{marder2019smile}. It has been observed that use of emojis by university staff in emails sent to students positively affects students' task behaviour \cite{marder2019smile}. Emotions of email texts have been analysed in various studies to understand the mental state of senders \cite{couss2009attr, shao2019analytical, shen2013understanding}. However, to the best of our knowledge, we did not observe any existing work that focuses on adding emojis to emails. Hence, we propose \textit{EmoG}, as a Gmail plugin to demonstrate the idea of augmenting emails with emojis, based on the sentiments being expressed in the text. The sentiment categories considered and emojis corresponding to these sentiments are presented in Figure \ref{fig:classemo}.

The remainder of this paper is structured as follows.
Section \ref{related work} discusses the related work , while Section \ref{design}, which focuses on design methodology and development of \textit{EmoG}. Working of \textit{EmoG} is presented in Section \ref{working} and Section \ref{user} describes user scenario. Evaluation and Results are presented in Section \ref{eval} Section \ref{results} respectively. Finally, we discuss the limitations in Section \ref{disc} and conclude the paper with future directions of enhancing \textit{EmoG} in the Section \ref{conclusion}.

\section{Related Work}
\label{related work}
\begin{figure*}[]
  \centering
    \includegraphics[width = \linewidth]{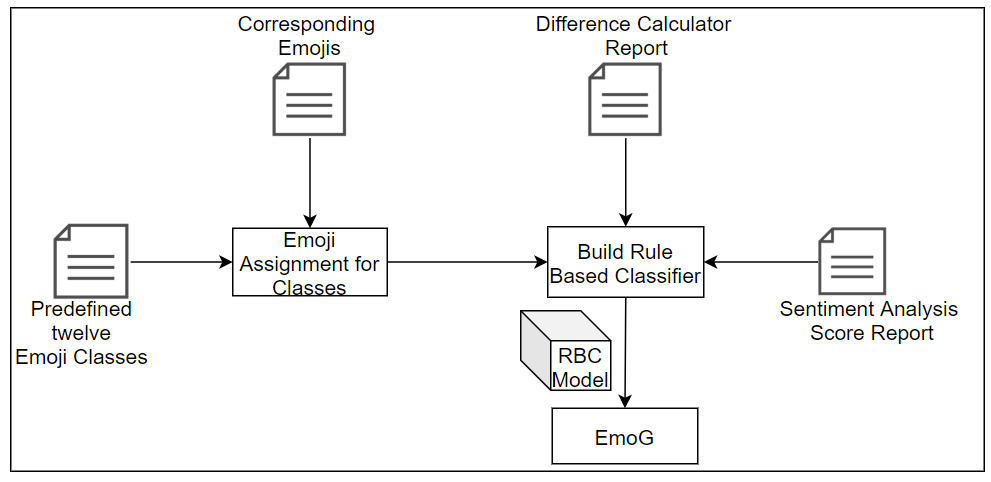}
    \caption{Design Methodology of \textit{EmoG}}
    \label{fig:approach}
\end{figure*}
Considering the extensive use of emails across various domains, several ideas have been proposed to enhance the experience of email users \cite{szostek2011dealing, sobotta2015capacity, sobotta2016mail, sobotta2016measuring}. 
A study has been conducted with 16 volunteers belonging to industry and academia to understand the needs with respect to handling emails. Six such needs have been identified through this study, which include sorting and annotating emails, reliable structure of the email with improved search and an clear overview of the email. These needs aim towards improving both the retrieval mechanism and the organization of emails \cite{szostek2011dealing}. Various items of emails including complex content, threaded conversations, structure of emails and so on have been identified to induce cognitive load on readers, in a study by Sobotta \cite{sobotta2016mail}. Need of improving emails to reduce cognitive load has been highlighted in this study \cite{sobotta2016mail}. Other studies also focused on various factors of emails that induce overload on readers with respect to extensive information and causes of information overload. These studies indicated the need for improvement of emails and email management strategies such that they are more user friendly \cite{sobotta2015capacity, sobotta2016measuring}. 


Park et al. have analysed user requirements for automatic email handling through a survey and a design workshop, that were aimed to understand the categories of automatic email handling requirements and the information required to address these needs \cite{park2019opportunities}. Based on the results of the survey and the design workshop, they have also proposed \textit{YouPS}, a programmable email system, that facilitates users to define custom email rules that could be involved in handling emails \cite{park2019opportunities}.
A study on response rates towards emails on web survey and face-to-face interview with respect to subject lines in the email revealed that many readers tend to ignore some parts of the emails in general \cite{sappleton2016email}. It has also been observed that readers prefer viewing emails with provocative subject lines than those with plain information \cite{sappleton2016email}. Adding sentiment of the email as a part of subject line might act as a provocative visual cue to the readers.
Email sentiments have also been explored in the literature. Coussement et al. have tried to classify emotions of emails sent by customers in a client-company environment into positive and negative categories using ML techniques \cite{couss2009attr}. They have compared three ML techniques and observed that Random Forest technique displayed better classification for client-email data in comparison to the considered Support Vector Machine and Logistic Regression algorithms \cite{couss2009attr}. Shen et al. have attempted to  predict personality of email writers by assigning scores on a 3-point scale (low, medium, high), for four personality dimensions presented in psychology \cite{shen2013understanding}. They have monitored around 49K emails in Outlook and 65K emails from Gmail inbox, and reported the number of subjects with low, medium and high personalities respectively \cite{shen2013understanding}. Abbas et al. have assessed emotions being conveyed in  emails exchanged among students and teachers in an academic environment, by considering Ekman's six emotion categories as a basis \cite{abbas2019sentiment}. Lanovaz et al. have attempted to classify emails related to R language into positive and negative categories to understand the nature of responses based on tone of the email, and observed a correlation between nature of response and tone of emails\cite{lanovaz2019comparing}. This re-emphasizes the influence on response based on sentiments in emails. An analytical system has been proposed to analyse emotions in email threads and consequently understand personality and mental state of senders \cite{shao2019analytical}. It also ranks the emotional personality of sender, on a scale of 0 to 1, indicating most emotional and least emotional personalities. A tweleve state emotion classification has been defined, which was observed to provide better insights than the existing emotion categories for the email dataset considered \cite{shao2019analytical}. 

Thus, understanding sentiments of emails could help in prioritizing emails and also might reduce cognitive load and stress on the readers by providing an emotional gist of the email. Ernst et al. have emphasized that use of emojis in emails could help in better email perception \cite{ernst2018effects}. Adding emojis in the subject line could also help in sorting emails in the inbox based on sentiments of the emails and also motivate readers to view the email. But, it is time taking to manually add emojis to emails using normal keyboard as it requires special keys. Though emoticons\footnote{emoticons include symbols such as :) , :( and so on, where as emojis refer to expressions shown on smiley faces} could be used to express sentiments using the keyboard keys, emojis could be more appealing to the readers than emoticons. Augmenting emails with emojis automatically at the receiver end might help in reducing effort of adding emojis to emails, to gain better expressiveness of sentiments. This also helps readers to take decisions on which parts of email are to be considered more prominently, when they do not intend to read the email completely. Though the existing literature emphasizes the need for analysing emotions in emails and also the importance of emojis in emails, we are not aware of any work that deals with automatically adding emojis as visual cues to emails. Hence, we propose \textit{EmoG}, as a Google Chrome Extension 
that can be added as Gmail plugin to append emojis to text in emails present in the inbox. 
\section{Design Methodology}
\label{design}
\textit{EmoG} has been designed based on the observation by researchers that email annotation is one of the six desired needs of email users \cite{szostek2011dealing}. Thus, a preliminary prototype version of \textit{EmoG}, to demonstrate the idea of annotating emails with emojis based on the sentiments being conveyed in emails has been developed. The design methodology of \textit{EmoG} is presented in Figure \ref{fig:approach}. 

There have been multiple studies in the literature that consider different number of emotional categories, from only two emotional categories - positive and negative\cite{lanovaz2019comparing} to more than ten emotional categories \cite{shao2019analytical}. Ekman's six emotion categories are among the widely used classifications \cite{ekman1992argument}. It classifies emotions into six categories - Happy, Sad, Fear, Disgust, Anger and Surprise \cite{ekman1992argument}. Another frequently used emotion categorization is classifying emotions into four categories - Anger, Anxiety, Positivity and Sadness \cite{shao2019analytical}. Ten emotion categories have been highlighted in \cite{russell1987relativity} during the evaluation of accuracy of automated face expression detectors. A different combination of four emotion categories - Joy, Sadness, Anger and Disgust have been used in a study towards identifying emotions in Arab tweets \cite{hussien2016emoticons}. Bao et al. have identified eight emotion categories through LDA topic modelling and manual inspection, in texts extracted from two different News channels \cite{bao2011mining}. Shao et al. have identified twelve emotional categories in emails of 41 employees of an organization. They have surveyed the literature for different emotional categorization and built upon the existing literature, predominantly on Ekman's six emotional categories, that are widely used for several studies. Several other combinations of emotion categories, with varied number of categories have been explored in the literature. Shao et al. have also compared the the 12 emotion categorization with existing four emotion categorization - Anger, Anxiety, Positivity and Sadness, and Ekman's six emotion categorization. This comparison revealed that the twelve emotion-category approach provides better insights with respect to emails, than other existing emotional classifications in the literature. Considering the relevance of twelve emotional categories to emails and the positive observations of this classification with respect to other existing classifications in the literature, we designed \textit{EmoG} based on the twelve emotional categories defined by Shao et al. in \cite{shao2019analytical}.

\begin{figure*}[]
  \centering
    \includegraphics[width = \linewidth]{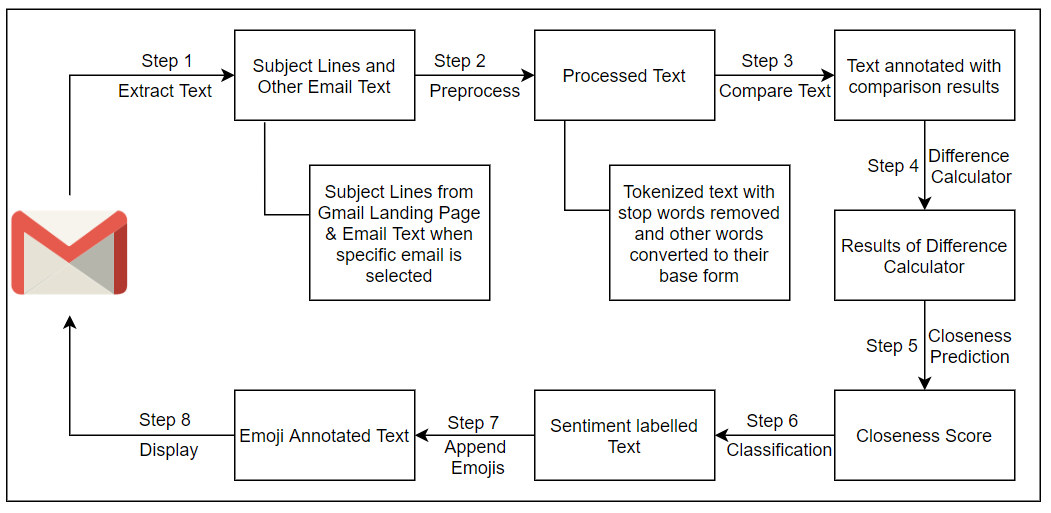}
    \caption{Working of \textit{EmoG}}
    \label{fig:working}
\end{figure*}

Considering these twelve categories, we created a corpus having words related to each emotion. Emotion related words are considered as keywords and synonyms of these keywords have been generated by running scripts that used NLTK corpus-based library, \textit{wordnet}. \textit{Wordnet} being a readily available, easy-to-use, synonym generator library, we could conveniently extract synonyms of keywords for each of the 12 emotions. The generated synonyms have been added to the corresponding keyword list, and the process of extracting synonyms is re-iterated for each of the new keywords in the list, until no new words are identified for each of these twelve categories. All the generated synonyms for each emotion category are aggregated to the corresponding keyword list and each class is assigned a numeric ID. 

A list of emojis that correspond to each of the 12 emotion classes have been identified. The leading keyword for each emotion category were used as search terms in an online emoji suggester and the first suggested emoji is considered to represent each of the twelve emotion categories. These emojis are then assigned to corresponding classes. A Rule Based Classifier Model has been built to classify text into the twelve emotion categories and append corresponding emoji to the text. This Rule Based Classifier is designed to identify the extent of similarity of a textual sentence with the all the keywords formulated for each of the emotional classes through a difference calculator. Based on the level of similarity, the textual sentence is classified into the class with highest similarity. The Rule Based Classifier also consists list of emojis to be appended for a specific emoji class, which is fetched and corresponding emoji is added to the textual sentence. 
This evaluation of similarity could also be considered as a closeness metric, where the value returned to the Rule Based Classifier by difference calculator reveals how close a given statement is, to each of the classes. The Rule Based Classifier is built based on the closeness metric score function to classify the statement into the closest class. The list of 12 classes considered and the assigned corresponding emojis are depicted in Figure \ref{fig:classemo}.

\section{Working of \textit{EmoG}}
\label{working}
\textit{EmoG} appends emojis to email text when added as a plugin to Google Chrome browser, through an eight step process as presented in Figure \ref{fig:working}.
This process is discussed below.

\begin{itemize}
    \item \textbf{Step 1: Extract Text-} As a first step, textual data displayed on the Gmail page for each email is extracted. The text in subject line of the emails in users' inbox is extracted, and when the user navigates to a specific email, text of the email is extracted.
    \item \textbf{Step 2: Text Preprocessing-} Text extracted is filtered to eliminate any special characters and stopwords that might be present, using the English stop-word removal methods provided by NLTK stemmer library. The filtered text is processed by NLTK stemmer library to generate base forms of each of the words in the text, resulting in a processed text data.
    \item \textbf{Step 3: Compare Text-} The processed text is compared with keywords in each of the emotion categories. Number of words in each textual sentence that are similar to keywords in each emotion category are identified and the textual sentence is annotated with these comparison results. The results include the number of words belonging to each of the twelve emotion categories.
    \item \textbf{Step 4: Difference Calculation-} The text annotated with comparison results is analysed and difference of number of words in a textual sentence with respect to similar words is calculated for each category. The formula presented below is used to calculate difference of a textual sentence with respect to each emotion category.
    \[D_{E_a} = T_{a} - SE_{a}\]
    where, D\textsubscript{E\textsubscript{a}} refers to Difference Score of a textual sentence with respect to Emotion category E\textsubscript{a}, T\textsubscript{a} refers to total number of words in the textual sentence and SE\textsubscript{a} refers to synonym keywords of emotion category E\textsubscript{a}. The results obtained by calculating difference based on this formula are stored for further processing.
    
    \item \textbf{Step 5: Closeness Predictor-} Difference calculator function evaluates the closeness of the preprocessed text with each of the 12 emotion classes, by comparing the text with list of keywords formulated for each of the classes. The class which matches the most to a sentence is identified by the majority number of words of the textual sentence that match with the keywords of the class. Thus, a closeness metric score with each of the classes is obtained. Closeness metric is defined as the inverse of value returned by difference predictor and is calculated as follows.
    \[C_{E_{a}} = (D_{E_{a}})^{-1} \]

    \item \textbf{Step 6: Classification-} A Rule Based Classifier classifies the extracted text into one of the twelve emotion classes based on the closeness value returned in the previous step. Lesser the difference between synonyms of words of the textual sentence and the keywords of emotion class, closer the textual sentence to the class. The text extracted is classified into the emotion class with the highest closeness metric score. It is derived based on the following formula.
    \[Emotion\_Category = max(C_{E_{a}}), \forall E\textsubscript{a}\]
    Based on this classification mechanism, the textual sentences are labelled with corresponding emotion class names.
    
    \item \textbf{Step 7: Append Emojis-} The emoji that corresponds to the class into which the extracted text is classified is added to the subject line and to the email text. The resultant text annotated with Emojis is stored.
    
    \item \textbf{Step 8: Display-} The emoji annotated texts obtained from the previous step are extracted and displayed over the corresponding textual sentences on the Gmail interface.
    
\end{itemize}

\section{User Scenario}
\label{user}
\begin{figure*}[]
   \centering
    \includegraphics[width = \linewidth]{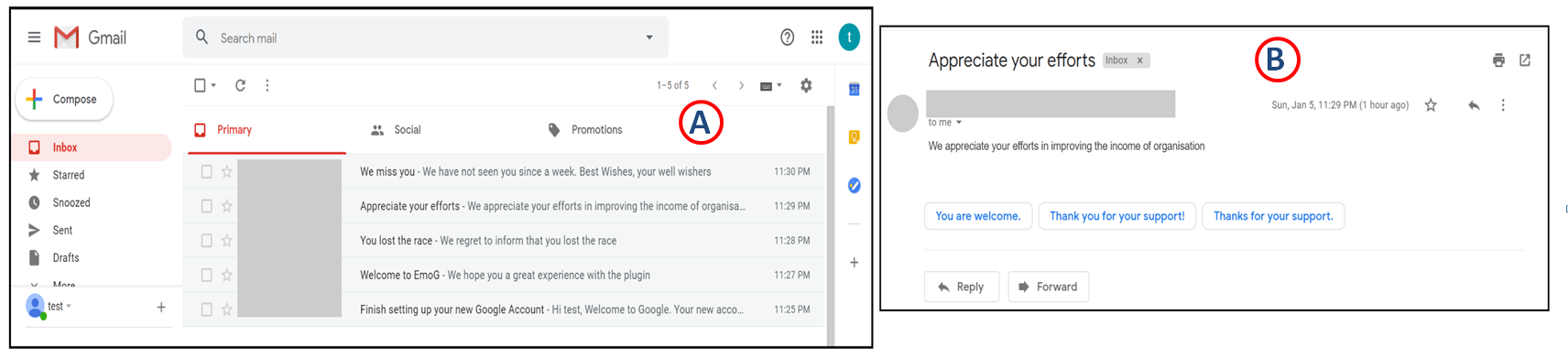}
    \caption{Snapshot of Gmail inbox landing page without \textit{EmoG}}
    \label{fig:without_emo}
\end{figure*}

Consider \textit{Veda} to be a student Gmail user who visits Gmail on her browser to check emails she might have received. \textit{Veda} sees that she has received many emails as shown in [A] of Figure \ref{fig:without_emo}, but observes she is running late to a class and hence she cannot read all the emails. She decides to read only a few important or exciting emails due to lack of time.

To do so, she reads through the subjects displayed on Gmail page. If she finds any email subject to be exciting or important, she opens the mail. She finds email body similar to the body shown in [B] of Figure \ref{fig:without_emo}, and considering the time constraint, she prefers to read only few points in the email. As a result, she tends to skip most of the email text and generally reads top few sentences of the email.
\begin{figure*}[]
  \centering
    \includegraphics[width = \linewidth, height = 6cm]{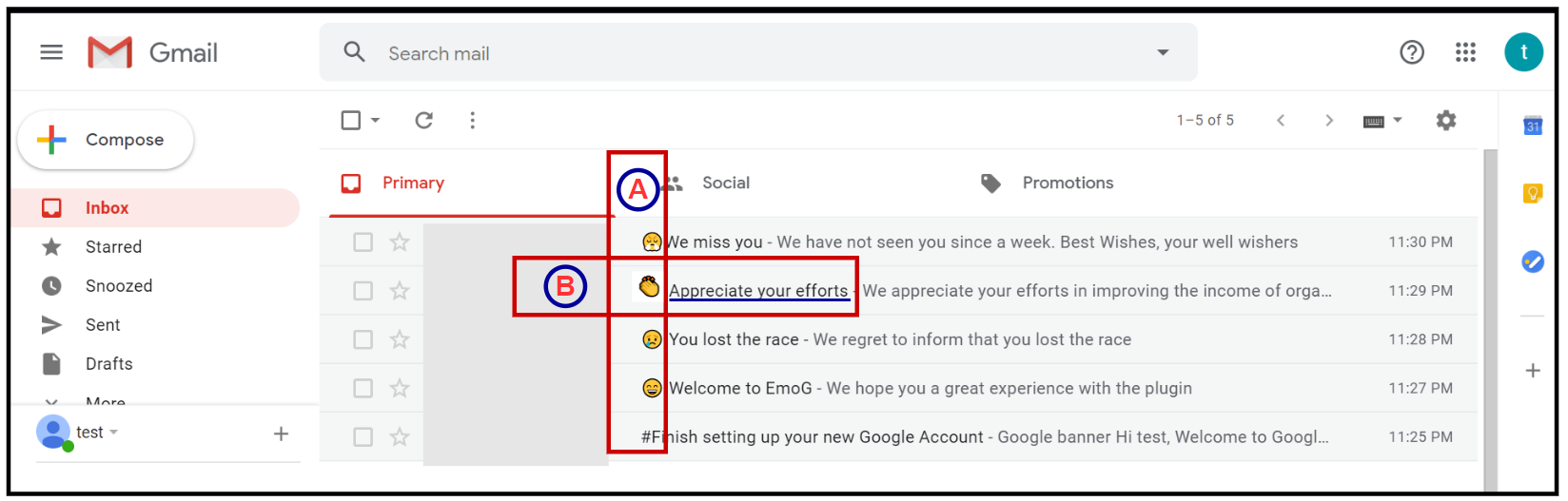}
    \caption{Snapshot of Gmail with \textit{EmoG}}
    \label{fig:user scenario}
\end{figure*}

\begin{figure}
    \centering
    \includegraphics[width= \linewidth]{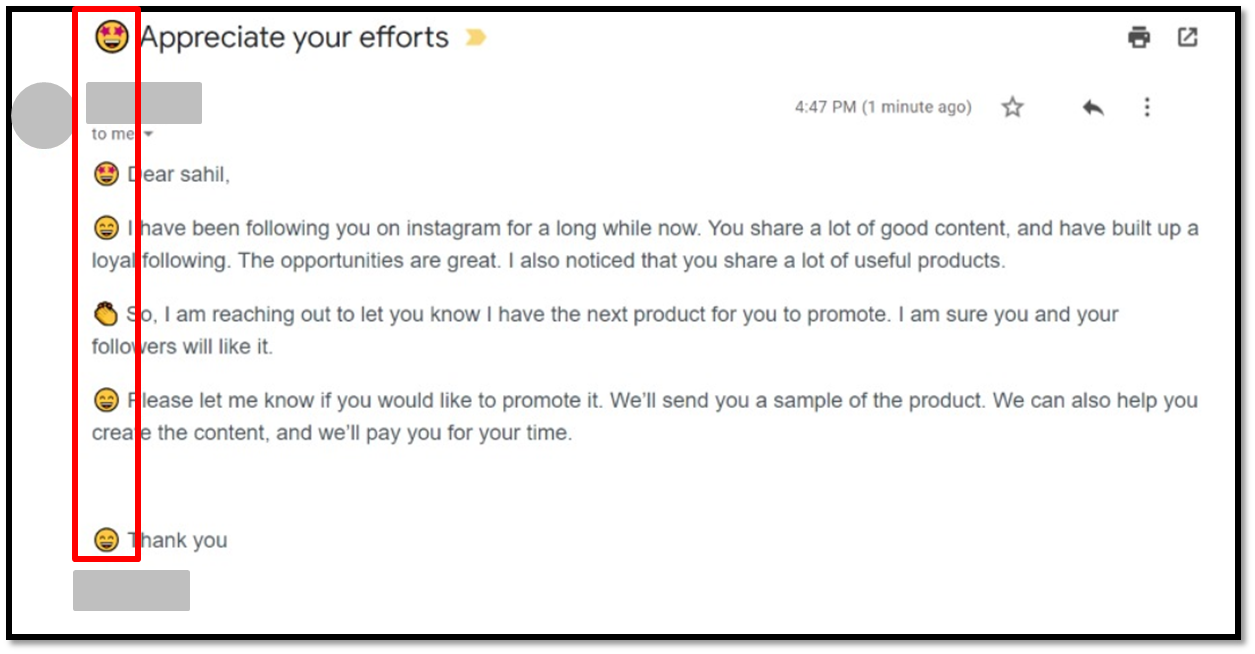}
    \caption{Snapshot displaying example of emoji added when navigated to email body\textsuperscript{2}}
    \label{fig:example_snap}
\end{figure}
She finds it time consuming to read through subjects of every email on Gmail homepage. She also wishes to read important and specific statements in the mail rather than going through the entire mail. She then installs \textit{EmoG} as a plugin to Google Chrome and reloads the Gmail page. She is now displayed with email subjects that are appended with respective emojis on the Gmail homepage as shown in [A] of Figure \ref{fig:user scenario}. \textit{Veda} glances through the page and decides on specific email based on the emoji that depicts the emotion that is expected to be conveyed in the email. Once \textit{Veda} navigates to body of the email, the text in the email is augmented with emojis based on the sentiment of the text as shown in Figure \ref{fig:example_snap}. \textit{Veda} now glances the page and reads only those points that are appended with emojis that she considers to be important.  
As she was preferring to read emails that are exciting and interesting, she scouts the Gmail inbox for subject lines annotated with emojis corresponding to Class 1, 2, 11 and 12, i.e., \textit{Glad, Praise, Good} and \textit{Interest} emotion classes, presented in Figure \ref{fig:classemo}. She hence selects email that is indicated by [B] of Figure \ref{fig:user scenario}, as it is augmented with \textit{clapping hands}, which refer to \textit{Praise} emotion category (Class 2 in Figure \ref{fig:classemo}). She then clicks on this email and navigates to the email page, which consists of emoji appended text as shown in Figure \ref{fig:example_snap}. Emails with larger texts are appended with emojis for multiple textual sentences as well.

\section{Evaluation}
\label{eval}
The main aim of \textit{EmoG} is to facilitate better motivated and stress free usage of emails. Hence, we evaluated \textit{EmoG} to understand the usefulness and player experience dimensions through a user survey. Towards this, we invited university students in their under-graduation and post-graduation courses to participate in the survey. We sent out emails to around 25 students asking for participation, of which 15 students were willing to participate. Thus, we have evaluated \textit{EmoG} with fifteen university students in the age group of 18-23 years. Nine of these participants were pursuing their under graduation course and six of them were pursuing their post graduation course during the user survey.

A short questionnaire has been drafted considering the usefulness and player experience criteria to be evaluated. This questionnaire, presented in Table \ref{tab:table2} is designed to be answered on a five point Likert Scale, based on participant opinion, from Strongly Disagree to Strongly Agree. The questions Q1 and Q5 refer to player experience criteria and questions Q2, Q3 and Q4 refer to usefulness criteria of \textit{EmoG}.

All the interested volunteers were reached out through emails that contained documents mentioning steps to install \textit{EmoG}, source of downloading \textit{EmoG} and a brief overview of \textit{EmoG} functionalities. The participants were then requested to download and install \textit{EmoG} from the specified location, as an extension to Google Chrome web browser on their personal laptops. The installation guide and functionality specification documents were shared in the same email, with the volunteers. This eased the installation of plugin and smooth flow of the survey. The participants were then asked to login to their Gmail accounts and skim through the Gmail homepage. They were then requested to view the body of few emails of their choice. After the survey, all the volunteers were asked to answer a 5-point likert scale based questionnaire as shown in Table \ref{tab:table2}, with respect to their experience during the survey.

Volunteers were informed prior to the survey that their email data is only being processed online by \textit{EmoG} and that no email data of the volunteers is stored for any further processing.

We have also evaluated correctness of \textit{EmoG} by manually verifying random email subject lines and textual sentences appended with emojis. We manually verified 105 textual sentences, of which 65 were subject lines and the rest 40 were texts in emails. We read through these sentences and tried to label the sentiment being conveyed by the specific sentence. Then we compared our labels to those labelled by \textit{EmoG}. 
\begin{table}
    \centering
    \begin{tabular}{|l|}
     \hline
     \textbf{Q1:} It was easy to install the plugin \textit{EmoG} \\(5=strongly agree, 1=strongly disagree)\\
   \\
   
     \textbf{Q2:} According to you, \textit{\textit{EmoG}}
     has rendered emotion\\ of posts satisfactorily, with appropriate emojis. \\(5=strongly agree, 1=strongly disagree)\\
   \\
     \textbf{Q3:} \textit{\textit{EmoG}} has helped me in
     identifying useful\\emails and
     email content through emojis\\
     (5=strongly agree, 1=strongly disagree)\\
    \\
 
      \textbf{Q4:} \textit{\textit{EmoG}} has motivated me
      to view emails, which \\helped
      in getting better insights about 
      emotional \\context of the 
      email. \\(5=strongly agree,
      1=strongly disagree)\\
    \\
    \textbf{Q5:} I will recommend \textit{\textit{EmoG}} 
    to my peers.\\ (5=strongly agree, 1=strongly disagree)\\
     \hline
 \end{tabular}
    \caption{Questions in survey using a 5-point Likert Scale}
    \label{tab:table2}
  \end{table}
  
\section{Results}
\label{results}
\begin{figure}
    \centering
    \includegraphics[width = \linewidth]{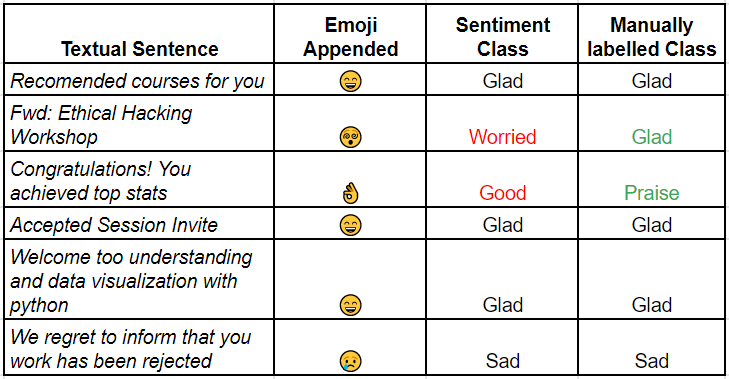}
    \caption{A brief list of randomly selected Email subject texts that have been used for manual evaluation.}
    \label{fig:man_res}
\end{figure}
\begin{figure}
    \centering
    \includegraphics[width= \linewidth]{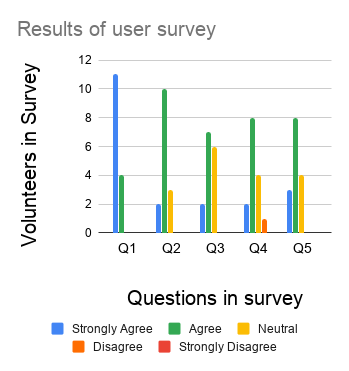}
    \caption{Results of user survey}
    \label{fig:results}
\end{figure}

A list of randomly selected six subject lines used for manual evaluation is presented in Figure \ref{fig:man_res}. It presents textual sentence used for evaluation, the emoji appended by \textit{EmoG}, its corresponding emotion class identified by RBC defined during development of \textit{EmoG}, emotion category identified through manual analysis.

The Sentiment Class of textual sentences that were misclassified is presented in \textit{red} color and the corresponding appropriate classification is presented in the adjacent column, with classes presented in \textit{green} color. The second and third textual sentences were recognised by \textit{EmoG} to express \textit{Worried} and \textit{Good} emotions. The second sentence indicates an email that describes \textit{Ethical Hacking Workshop}, which could be classified as \textit{Glad}. However, it has been classified as \textit{Worried}. We assume that this classification is due to the presence of \textit{Hack} word in the sentence, which generally implies a negative emotion in majority of the cases. Also, the third statement is classified as \textit{Good}, while it could be classified as \textit{Praise} as it \textit{Congratulates} the receiver for their achievements. This misclassification might be due to the almost similar keywords that could be present in \textit{Good} and \textit{Praise} classes. The manual analysis of all the 105 textual statements resulted in an overall accuracy of 72.4\% in terms of emotion classification. We observed that 50(80\%) subject lines and 26(65\%) email texts were correctly labelled according to the emotions being conveyed. We assume that the misclassification is due to the consideration of keyword-comparison. The misclassifaction might also be due to 
similar keywords in almost close emotion categories. This comparison masks the intent of the statement and is dependant solely on the words in the sentence. Using ML and superior NLP techniques to consider the intent of sentence might help in improving accuracy of the classification.

We evaluated the user survey responses based on Likert scale points, shown in Table \ref{tab:table2}, for each question. The percentage of responses for each of the weights is considered as the evaluation metric for individual question and a cummulative of all the percentages is considered as an evaluation metric for \textit{EmoG}. The score for each question is the percentage of ratio sum of points obtained on the Likert Scale to the maximum possible points (here 75 for each question, considering strongly agree to be marked by all 15 participants for the question). The value of points ranges from 5 to 1 implying \textit{Strongly Agree} to \textit{Strongly Disagree} . As reported in Figure \ref{fig:results}, it is observed that \textit{\textit{EmoG}} had a good user-friendly interface (94\% in Q1). In Q2, about 80\% of participants have either agreed or strongly agreed that \textit{\textit{EmoG}} has satisfactorily rendered emotions to emails, with appropriate emojis (score of 78.6\% in Q2).
The scores in Q3 and Q4 infer that \textit{\textit{EmoG}} has helped volunteers to identify useful comments and motivated them to view emails and get better insights (scores: 74.6\% in both Q3 and Q4). Participants have mentioned their suggestions to enhance \textit{\textit{EmoG}} by reconfiguring the emojis. In Q5, 11 of 15 participants have either agreed or strongly agreed to recommend \textit{\textit{EmoG}} to their peers and rest of them replied with a neutral feedback(score:78.6\%). Participants have also pointed out that -
\textit{emojis are being wrongly appended to few emails} and \textit{navigating to a new mail requires refreshing the whole page, which is time taking sometimes}

\section{Discussion and Limitations}
\label{disc}
\textit{EmoG} has been developed as a Gmail plugin to support stress-free and motivated email management. It appends one of the 12 emojis to text in Gmail based on the sentiments expressed in the sentences.

Currently, the identification of sentiment of textual sentences is purely based on Rule Based Classifier model defined during development of \textit{EmoG}. This model depends on the synonym keywords for each of the twelve emotions and the similarity calculator. It has also been observed that some texts are not labelled correctly. The dependencies and the techniques used for comparison have a significant effect on the accuracy of classification. A more reliable classification model using Machine Learning, such as Support Vector Machine or Random Forest Generator could be implemented to enhance accuracy. Also, the current labelling of emotions is dependant on comparison of synonyms. This could be replaced with superior Natural Language Processing(NLP) techniques such as Latent Dirichlet Allocation (LDA), for better identification emotions in textual sentences. The twelve emotion categories considered for appending emojis to emails are based on existing literature presented by Shao et al \cite{shao2019analytical}. We considered this set of classification as it is more relevant when compared to other existing classifications in the literature and deals with similar dataset, i.e., Emails. However, the relevance of other classifications could be explored and \textit{Neutral} emotion category could also be added, as it is important to include emails with neutral sentiments. Also, a qualitative survey could be conducted to understand the frequent emotion categories observed in emails and to narrow down on to specific emotion taxonomy in Emails, based on the insights from existing literature and qualitative survey.

The existing user survey included participants only from academia, though the plugin could be universally used by Gmail users in other fields. An extensive survey could be conducted using better evaluation approaches and include participants from varied backgrounds and wider age groups, that include individuals from software industry, marketing industry, career guidance and counselling firms and so on to understand preference of email users with respect to annotation of emails with emojis. While emojis might add value to emails, there could be some emails, predominantly in the business domain, that might not require to be appended with emojis. In the current state, \textit{EmoG} has been designed to support only Gmail application on Google Chrome browser. It can further be extended to support other browsers such as Mozilla Firefox and Safari and to support other email applications such as outlook, yahoo and so on. The manual analysis followed for evaluation is restricted only to 105 email textual sentences, and is specific to the statements considered. Accuracy levels of sentiment classification and augmentation of emojis depend on textual statements considered and might differ if a different set of textual sentences are considered.
\section{Conclusion and Future Work}
\label{conclusion}

In this paper, we presented an initial prototype version of \textit{EmoG}, as a Google Chrome plugin to append email content in Gmail with emojis, based on emotions being conveyed in the emails. Extensive use sometimes overloads users with emails. In such scenarios, users prefer to prioritize emails and address them. One possible prioritization metric could be based on the emotions being conveyed in the emails. However, it has been observed that it is difficult to express emotions through pure text. Adding emojis to text might convey emotions in a better way. However, adding emojis while drafting emails is not a common practice and it is also time-consuming as the emojis are not directly available as keys on conventional keyboards of laptops or computers. \textit{EmoG} appends emojis to the email content, based on the emotion conveyed in that text.
We have classified text in emails into one of the 12 emotion categories proposed in the literature, using a Rule Based Classifier. The text is appended with emoji corresponding to the classified emotion class, thus attempting to provide insights on emotions involved in emails. It might also support decision making in prioritizing emails and in sorting emails, considering the appended emojis as tags. Since \textit{EmoG} is intended to support individuals in an educational institution, specifically targeted towards students, adding visual cues by appending emojis might motivate students to read the emails. We conducted the user survey with 15 university students using a 5-point Likert scale and obtained promising results, with 80\% of the participants willing to suggest \textit{EmoG} to their peers.

\textit{EmoG} presented in the paper is a preliminary prototype version and hence is designed only to classify based on 12 classes defined in the literature \cite{shao2019analytical}. We plan to improve the classification based on emotion-classes integrated from multiple research works, and consequently develop a theoretical model of emotions in emails in future. \textit{EmoG} could be extended to support more emotions. We plan to improve the classification model being used to machine learning based approaches that use superior NLP techniques such as LDA, capable of automatically identifying topics in a textual sentence. This could overcome misclassification and consequently enhance the accuracy. We also plan to address the issue of repetitive page refreshes required while navigating back and forth from Gmail home page to Email page. Email senders might be given an option to send emails that are automatically annotated with emojis instead of the annotation being done at the receiver end. Users could be facilitated to correct the inaccurate emojis, which could  make \textit{EmoG} learn appropriate emoji-labelling over time, thus making \textit{EmoG} adaptable. \textit{EmoG} could be further designed as a personalised user-specific tool that learns from the user perceptions of emoji meanings and appends emojis accordingly, overtime.  

As an extension to the evaluation study, we also plan to include participants from varied backgrounds, both from academia and industry. We anticipate that this evaluation could provide better insights to improve the plugin. Also, to decide on the number of emotion categories, we plan to conduct qualitative surveys that include one-on-one interviews with Email users from industry and from academia, thus, including opinions from developers, managers, faculty and students. Sometimes, different emojis convey similar expressions. We plan to identify such emojis and add them to corresponding emotion classes. This can be used to append multiple similar emojis to a single textual sentence, to improve motivation. We shall also explore the possibility of single sentence conveying multiple emotions and append emojis accordingly.

\section*{Acknowledgements}
We thank our under graduate student, Vagavolu Dheeraj for helping us in developing the plugin. Also, we thank all the participants for their valuable time and honest feedback during the user survey.

\bibliographystyle{ACM-Reference-Format}
\bibliography{sample-base}


\end{document}